\documentclass[preprint,secnumarabic,amssymb, superscriptaddress, nobibnotes, aps, prb]{revtex4-2}
\usepackage{amssymb,epsfig,graphicx,multirow,bm,dcolumn,adjustbox}

\begin{document}

\title{Randomly packed Ni$_2$MnIn and NiMn structural units in off stoichiometric Ni$_2$Mn$_{2-y}$In$_y$ alloys}
\author{R. Nevgi}
\affiliation{School of Physical and Applied Sciences, Goa University, Taleigao Plateau, Goa 403206 India}
\author{E. T. Dias}
\affiliation{School of Physical and Applied Sciences, Goa University, Taleigao Plateau, Goa 403206 India}
\author{K. R. Priolkar}
\email{krp@unigoa.ac.in}
\affiliation{School of Physical and Applied Sciences, Goa University, Taleigao Plateau, Goa 403206 India}

\date{\today}

\begin{abstract}
Ni$_2$Mn$_{2-y}$In$_y$ alloys transform from the martensitic $L1_0$ antiferromagnetic ground state near $y = 0$ to austenitic ferromagnetic $L2_1$ Heusler phase near $y = 1$ due to doping of In impurity for Mn. The off stoichiometric alloys prepared by rapid quenching are structurally metastable and dissociate into a mixture of $L2_1$ (Ni$_2$MnIn) and $L1_0$ (NiMn) phases upon temper annealing. Despite this structural disintegration, the martensitic transformation temperature remains invariant in the temper annealed alloys. Investigations of the local structure of the constituent atoms reveal the presence of strongly coupled Ni$_2$MnIn and NiMn structural units in the temper annealed as well as the rapidly quenched off stoichiometric Ni$_2$Mn$_{2-y}$In$_y$ alloys irrespective of their crystal structure. This random packing of the $L2_1$ and $L1_0$ structural units seems to be responsible for invariance of martensitic transition temperature in the temper annealed alloys as well as the absence of strain glass transition in rapidly quenched alloys.
\end{abstract}

\maketitle

\section{Introduction}

The interplay between the structural and the magnetic degrees of freedom arising out of martensitic transition in NiMn based magnetic shape memory alloys is the source of their functionalities. The alloys display a variety of effects such as the magnetic shape memory effect \cite{Webster491984}, magnetic superelasticity \cite{Sutou200485,Krenke200542,Krenke200673,Krenke200775,Kainuma2006957}, the magnetocaloric effect \cite{Krenke20054,Moya200775}, giant magnetoresistance \cite{Chatterjee200942}, exchange bias \cite{Khan200791}, kinetic arrest \cite{Ito200892,Sharma200776} etc. Below 973 K, in its martensitic state, the binary alloy, NiMn, displays a tetragonal $L1_0$ structure with an antiferromagnetic order \cite{Kren291968}. Systematic doping of Z (Z = In, Sn, Sb) to realize Ni$_2$Mn$_{2-y}$Z$_y$ results in the transition of $L1_0$ structure to $L2_1$ Heusler phase mediated by a modulated phase. The magnetic ground state converts from antiferromagnetic to ferromagnetic as the dopant concentration approaches $y$ = 1 \cite{Sutou200485,Planes200921}. The progressive addition of the Z element results in a decrease of martensitic transition temperature followed by its disappearance at a critical concentration depending upon the type of Z atom (In, Sn, or Sb) \cite{Aksoy200791,Siewert201214,Entel201365}. At intermediate doping, Ni$_2$Mn$_{2-y}$Z$_y$  alloys exhibit a modulated monoclinic structure described as incommensurate 5M or 7M in their martensitic state and a mixture of ferro and antiferromagnetic interactions  \cite{Righi200755,Righi200856}. The structural studies performed using extended x-ray absorption fine structure (EXAFS) spectroscopy have shown the  presence of local structural distortion in the form of a shorter Ni--Mn bond distance than Ni--In bond distance \cite{Bhobe200674,Bhobe200820,Lobo201096}. This structural distortion is considered to be the microscopic driving force for the martensitic transition.  The structural distortion also results in the hybridization of Ni $3d$ and the $3d$ states of Mn atom present at the Z site of X$_2$YZ Heusler structure. It is believed to be the cause of antiferromagnetic interactions in these alloys \cite{Ye2010104,Priolkar942011}. Competing ferromagnetic and antiferromagnetic interactions are responsible for the superparamagnetic like ground state \cite{Cong962010,Cong2512014} in these alloys. Impurity like Fe doping in Ni-Mn-In alloys induces a transition from the ferroelastic ground state to a strain glass \cite{Nevgi1122018}. The ferroelastic/martensitic to strain glass transition usually occurs due to the formation of point defects when the dopant concentration exceeds a critical value \cite{Sarkar952005}. Segregation of defect phases in such impurity-doped alloys destroy the long-range ordering of the elastic strain vector, driving the system to a non-ergodic ground state \cite{Nevgi1032021,Nevgi542021}. Despite doping an impurity like In for Mn in NiMn martensitic alloy, a transition to strain glassy state is not reported in Ni$_2$Mn$_{2-y}$Z$_y$ alloys.

Recent studies have shown these Ni$_2$Mn$_{2-y}$Z$_{y}$ (Z = Ga, In, Sn, Sb) alloys are structurally metastable. Upon temper annealing, they disintegrate into a dual-phase composite alloy consisting of Heusler $L2_1$ and tetragonal $L1_0$  phases \cite{Cakir62016,Krenke1202016,Cakir4482018,Wanjiku1252019}. The phenomenon of shell ferromagnetism is a result of such structural metastability and occurs when the annealing is performed in an external magnetic field. The structural decomposition of Ni$_2$Mn$_{2-y}$Z$_{y}$ ($0 < y < 1$) into the $L2_1$ (Ni$_2$MnZ) and $L1_0$ (NiMn) phases or supercells depends on the time and temperature of annealing \cite{Cakir1272017}. The alloys with lower content of Z element disintegrate at a lower temperature and less time than those with higher Z content leading to time-dependent effects such as strong atomic relaxation. The phase separation due to decomposition is reported to be the lowest in energy and highlights the impact of structural disorder and the segregation of alloys close to martensitic transformation \cite{Entel2252018}. Therefore, are the local structural distortions observed through EXAFS even in the cubic phase of the off stoichiometric Ni$_2$Mn$_{2-y}$Z$_y$ alloys, precursors to the structural decomposition seen on temper annealing? Further, despite local structural disorder and metastable crystal structure, absence of a non-ergodic state like the strain glass remains intriguing.

The $L2_1$ structure of Ni$_2$MnZ (Z = Sn, In or Sb) with $a\sim$ 6.0 \AA~  and the $L1_0$ unit cell of NiMn ($a =$ 3.7 \AA, $c =$ 3.5 \AA) have  significantly different near neighbor distances. Therefore, it should be possible to understand the structural disintegration of the off stoichiometric Ni$_2$Mn$_{2-y}$Z$_y$ alloys into Ni$_2$MnZ and NiMn phases. It is also equally crucial to understand the role of structural distortions present in the off stoichiometric alloys in facilitating their dissociation into the two end members with $L2_1$ and $L1_0$ structures. To understand these aspects, we have studied, and compared local structures of Ni and Mn in Ni$_2$Mn$_{2-y}$In$_{y}$  $0 \leq y \leq 1$ alloys, prepared via rapid quenching with their temper annealed counterparts. All the alloy compositions were given the same temper annealing treatment (427 $^\circ$C for $\sim 10^5$ seconds) to understand the process of disintegration and the associated changes in their structure and properties. A detailed structural, thermal, magnetic, and local structural study shows that all the rapidly quenched off stoichiometric compositions, Ni$_2$Mn$_{2-y}$In$_y$ consist of $L2_1$ and $L1_0$ structural units tend to segregate and phase separate upon temper annealing.

\section{Experimental techniques}

High purity  elements (purity $\ge$ 99.99\%) Ni, Mn and In were used in the preparation of alloys by arc melting in an argon atmosphere. The homogeneity of the ingot was ensured by flipping the individual alloy multiple times. The prepared ingots were cut, powdered and covered in tantalum foil to be vacuum sealed in quartz tubes and annealed at 750 $^\circ$C. On completion of 48 hours, the samples were ice quenched to obtain the rapid quenched (RQ) alloy series. The compositions verified using scanning electron microscopy with energy dispersive x-ray (SEM-EDX) technique were within 2\% of the targeted values. A part of such  alloy compositions was further annealed at 427 $^\circ$C for 28 hours and allowed to furnace cool to room temperature to achieve temper annealed (TA) alloy series. Both RQ and TA series were characterized using the structural, thermal, magnetic and local structural probes. LeBail analysis of room temperature x-ray diffraction patterns recorded on a X'Pert diffractometer using Cu K$\alpha$ radiation was carried out using Jana 2006 software\cite{Petricek2292014}. Differential scanning calorimetry (DSC) measurements using a Shimadzu DSC-60 in the 200 K - 800 K temperature range were carried out on 8 mg pieces of each alloy crimped in an aluminum pan. Similarly, Differential thermal analyzer (DTA) measurements using Shimadzu DTG-60 were employed on small pieces ($\sim$ 8 mg) of each alloy placed in a platinum pan. Measurements were done by heating/cooling the sample (in inert atmosphere) from ambient temperature to 1100 K at a constant rate of 5 K/min.  The temperature dependent magnetization measurements M(T) were performed in the temperature range of 5 K -- 350 K. The samples were first cooled to 5 K in zero applied magnetic field. Data was then recorded while warming in 0.01 T applied field (zero field cooled (ZFC))  and in successive field cooled cooling (FCC) and field cooled warming (FCW) cycles.  Extended X-ray Absorption Fine structure (EXAFS) measurements at the P65 beamline, PETRA III Synchrotron Source (DESY, Hamburg, Germany) and BL-9C at Photon factory, KEK, Japan, were implemented to understand the local environment at Ni K (8333 eV) and Mn K (6539 eV) edges at room temperature. The absorbers were prepared by uniformly coating the alloy powders on scotch tape. The thickness of the absorbers was adjusted by controlling the number of layers of scotch tape so as to obtain the absorption edge jump $\Delta\mu t\leq 1$.  Here $\Delta\mu$ is the change in the absorption coefficient at the absorption edge, and $t$ is the thickness of the absorber. Using the gas ionization chambers as detectors, the incident (I$_0$) and the transmitted (I) photon energies were simultaneously recorded. At each K edge, at least three scans were collected to average the statistical noise. The data was analyzed using well established procedures in Demeter suite \cite{Raval200512}.

\section{Results}

\begin{figure}[h]
\begin{center}
\includegraphics[width=\columnwidth]{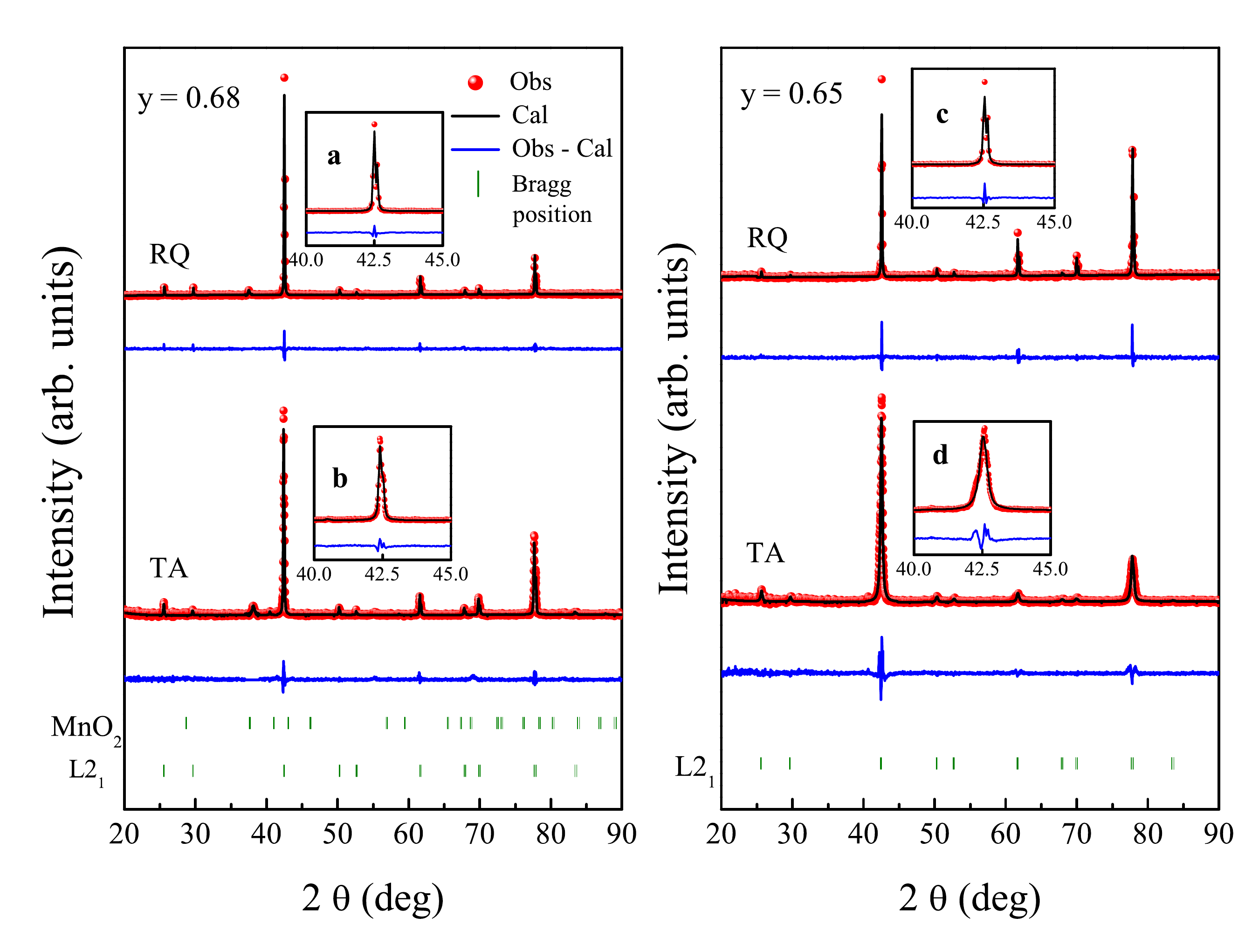}
\caption{X-ray diffraction data for the alloys $y$ = 0.68 and $y$ = 0.65. The broadening of major 220 peak is highlighted in rapid quenched (RQ) as inset (a) and temper annealed (TA) as inset (b) in $y$ = 0.68 and as insets (c) (RQ) and (d) (TA) in $y$ = 0.65.}
\label{fig:XRD1}
\end{center}
\end{figure}

\begin{figure}[h]
\begin{center}
\includegraphics[width=\columnwidth]{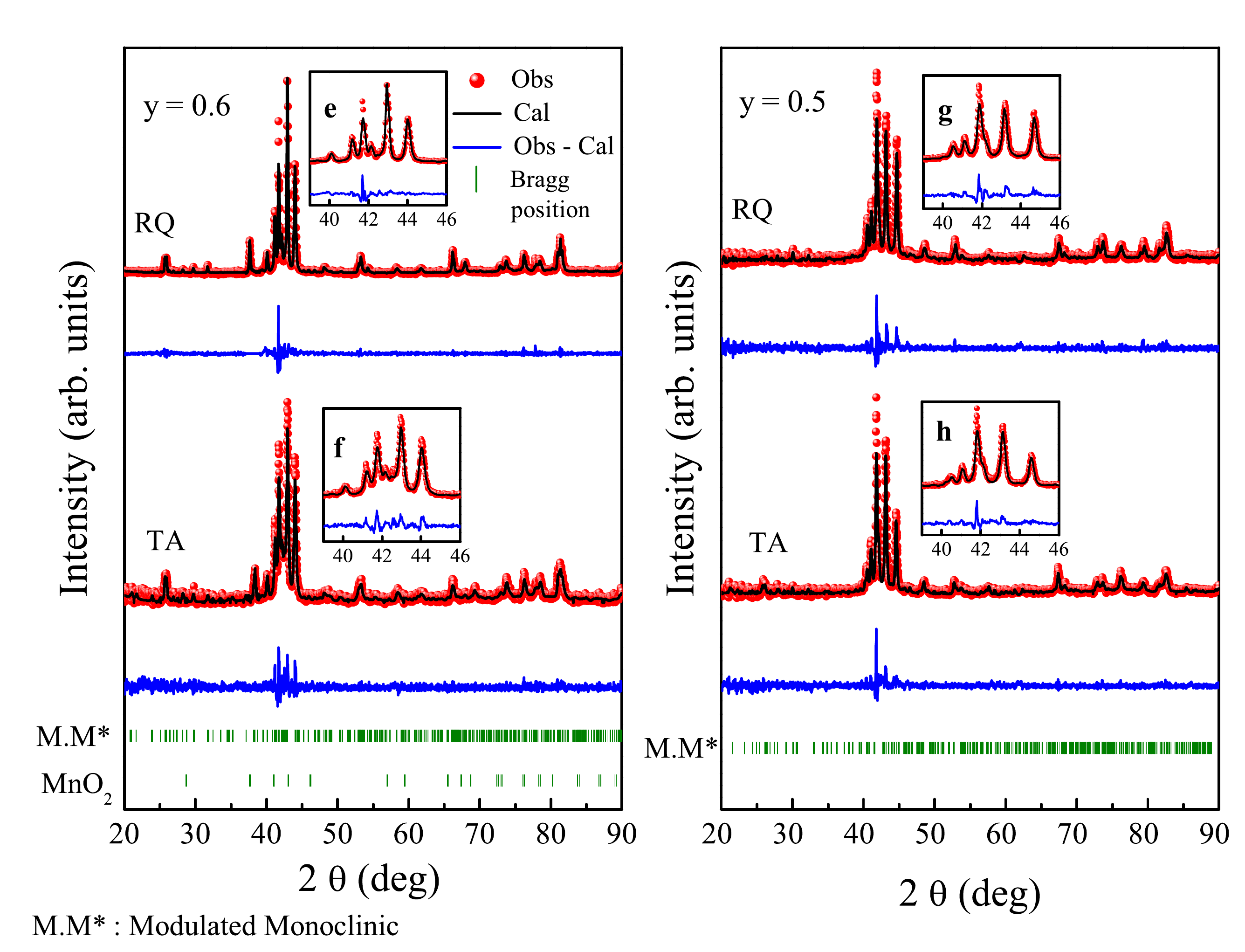}
\caption{X-ray diffraction data for the alloys $y$ = 0.6 and $y$ = 0.5. Insets (e), (f) and (g), (h) compare the change in the intensity of the major and satellite peaks due to a possible change in modulations or phase separation in rapidly quenched (RQ) and temper annealed (TA) alloys with $y$ = 0.6 and $y$ = 0.5 respectively.}
\label{fig:XRD2}
\end{center}
\end{figure}

\begin{figure}[h]
\begin{center}
\includegraphics[width=\columnwidth]{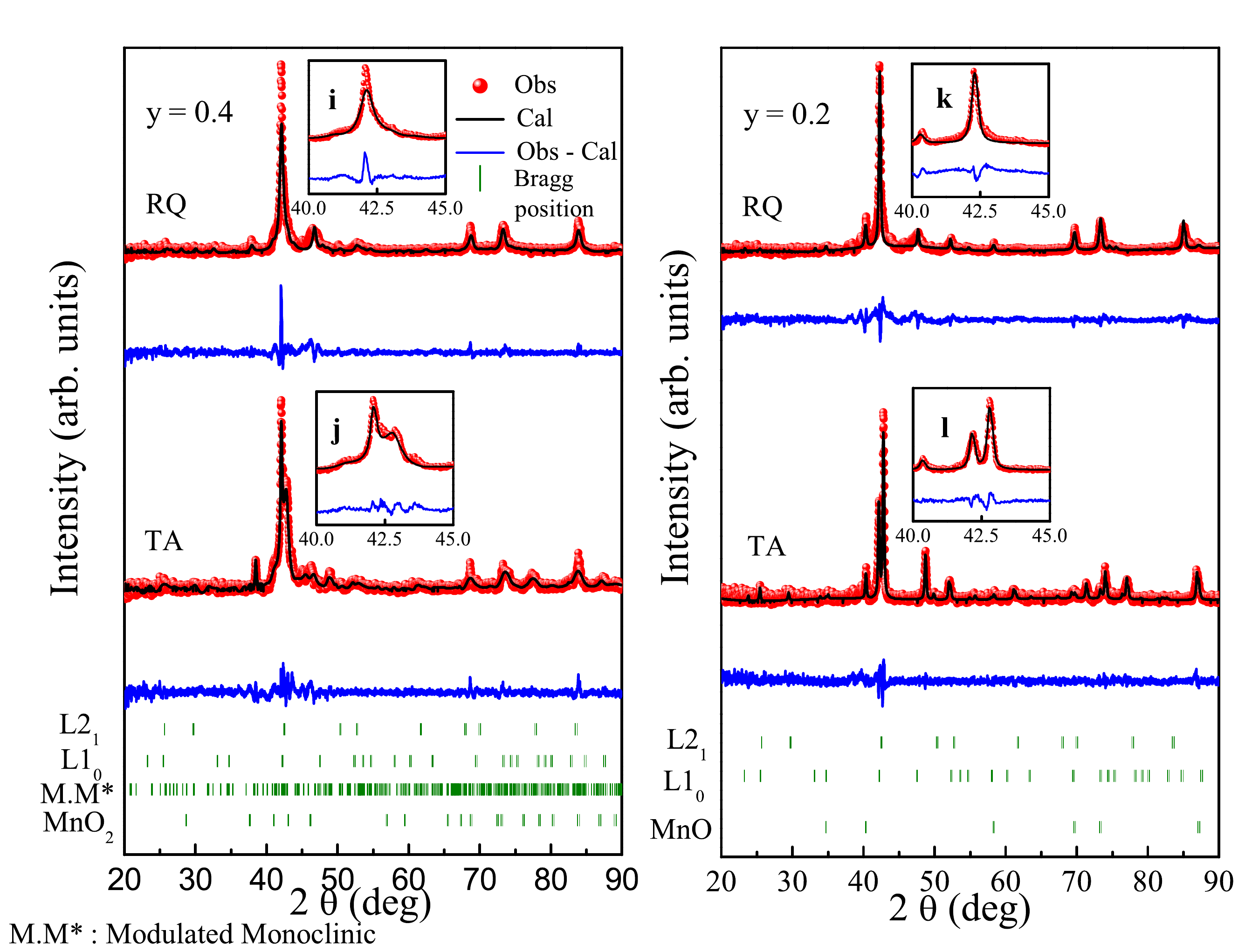}
\caption{X-ray diffraction data for the rapid quenched (RQ) and temper annealed (TA) alloys $y$ = 0.4 and $y$ = 0.2. The insets features the evolution of phases from $L1_0$ (i) to mixture of $L1_0$, $L2_1$ and modulated phases (j) in the alloy $y$ = 0.4 and the transformation of $L1_0$ (k) to mixture of $L1_0$ and $L2_1$ phases (l) in the alloy $y$ = 0.2}
\label{fig:XRD3}
\end{center}
\end{figure}

Room temperature x-ray diffraction patterns of the rapid quenched (RQ) and the temper annealed (TA) alloys along with the phases identified using the Le Bail method, are shown in the Figs. \ref{fig:XRD1}, \ref{fig:XRD2} and \ref{fig:XRD3}. Both RQ and TA alloys with $y$ = 0.68 and $y$ = 0.65 exhibit cubic Heusler phase as shown in Fig. \ref{fig:XRD1}. After temper annealing, even though  there is no apparent change in the structure, broadening of the diffraction peaks is noticed. Insets (a) and (b) for $y$ = 0.68 and insets (c) and (d) for $y$ = 0.65 clearly show the broadening of 220 Bragg reflection in TA alloys. The RQ alloys $y$ = 0.6 and $y$ = 0.5 presented in Fig. \ref{fig:XRD2} display a modulated martensitic structure, which appears to have been retained even after temper annealing. However, a difference in the intensity ratios between the main and satellite peaks is noticed in both the alloys $y$ = 0.6 (see insets (e) and (f)) and $y$ = 0.5 (see insets (g) and (h)) after temper annealing indicating a change in the modulation or partial disintegration of the alloys. It must be mentioned here that the structure of the martensitic state is debatable with different types of modulations being reported that depend on the alloy stoichiometry \cite{Krenke200673,Yan201588,Devi2018}. Interestingly, the tetragonal $L1_0$ structure of the RQ alloy $y$ = 0.4 converts into a mixture of $L1_0$, Heusler $L2_1$, and modulated phases upon temper annealing (Fig. \ref{fig:XRD3}, compare insets (i) and (j)). On the other hand, the alloy $y$ = 0.2 completely disintegrates  from a pure $L1_0$ phase into 22\% Ni$_2$MnIn and 72\% Ni$_2$Mn$_2$ along with a minor 6\% MnO phase (Fig. \ref{fig:XRD3}, inset (l))  after temper annealing. The crystallographic data obtained from the refinement are summarized in  Table \ref{table:table1}.

\begin{table*}[htbp]
\caption{The crystallographic data for the rapid quenched (RQ) and temper annealed (TA) alloys in the series Ni$_2$Mn$_{2-y}$In$_{y}$. $^\#$The shortfall in the sum of \% phase fractions is made up by impurity phases, MnO and MnO$_2$.}
\label{table:table1}
\begin{adjustbox}{width=\columnwidth,center}
\begin{tabular}{|l|l|l|l|l |}
\hline
\textbf{Compositions}      &\textbf{Type}  &\textbf{Phases (\% fraction)}    &\textbf{Space group}         &\textbf{Lattice parameters}  \\[0.5ex]
\hline\hline
\multirow{2}{*}{$y$ = 0.68}     &RQ$^\#$        &Cubic (99)               & Fm$\bar{3}$m                       & a = 6.01334(5) \AA  \\
\cline{2-5}
                                &TA$^\#$        &Cubic (98)              & Fm$\bar{3}$m                       & a = 6.01925(8) \AA  \\
\hline
\multirow{2}{*}{$y$ = 0.65}     &RQ        &Cubic (100)              & Fm$\bar{3}$m                       & a = 6.00701(5) \AA   \\
\cline{2-5}
                                &TA        &Cubic (100)              & Fm$\bar{3}$m                       & a = 6.0101(2) \AA   \\
\hline
\multirow{2}{*}{$y$ = 0.6}      &RQ         &Modulated monoclinic (100)             & I2/m($\alpha 0 \gamma$)00   & a = 4.3892(2) \AA, b = 5.6424(2) \AA, c = 4.3337(1) \AA\\
                                &           &                   &                             & $\beta$ = 92.91(4)$^\circ$, $q$ = 0.3321(4)$c^*$\\
\cline{2-5}
                                &TA$^\#$         &Modulated monoclinic (99)                & I2/m($\alpha 0 \gamma$)00   & a = 4.3842(2) \AA, b = 5.6414(4) \AA, c = 4.3291(3) \AA\\
                                &           &                   &                             & $\beta$ = 92.88(8)$^\circ$, $q$ = 0.3319(3)$c^*$ \\
\hline
\multirow{2}{*}{$y$ = 0.5}      &RQ         &Modulated monoclinic (100)                & I2/m($\alpha 0 \gamma$)00   & a = 4.4024(3) \AA, b = 5.5470(4) \AA, c = 4.3240(2) \AA\\
                                &           &                   &                             & $\beta$ = 94.226(6)$^\circ$, $q$ = 0.3101(3)$c^*$ \\
\cline{2-5}
                                &TA        &Modulated monoclinic (100)                 & I2/m($\alpha 0 \gamma$)00   & a = 4.405(3) \AA, b = 5.5538(4) \AA, c = 4.3281(3) \AA \\
                                &          &                    &                             & $\beta$ = 94.075(7)$^\circ$, $q$ = 0.3021(2)$c^*$ \\ [1ex]
\hline
\multirow{2}{*}{$y$ = 0.4}      &RQ       &Tetragonal (100)           & P4/mmm                      & a = 3.862(5) \AA, c = 3.468(9) \AA \\
\cline{2-5}
                                &TA$^\#$       &Tetragonal (54)           & P4/mmm                      & a = 3.736(1) \AA,  c = 3.514(1) \AA \\
                                &         &Cubic (11)                & Fm$\bar{3}$m                       & a = 6.0694(7) \AA \\
                                &         &Modulated monoclinic (33)                   & I2/m($\alpha 0 \gamma$)00   & a = 4.404(2) \AA, b = 5.628(2) \AA, c = 4.337(1) \AA \\
                                &          &                     &                             & $\beta$ = 94.17(2)$^\circ$ , $q$ = 0.3347(6)$c^*$  \\
\hline
\multirow{2}{*}{$y$ = 0.2}    &RQ      &Tetragonal (100)               & P4/mmm                       & a = 3.826(5) \AA,  c = 3.495(7) \AA \\
\cline{2-5}
                              &TA$^\#$       &Tetragonal (72)            & P4/mmm                       & a = 3.7364(2) \AA, c = 3.5135(4) \AA \\
                              &         &Cubic (22)                 & Fm$\bar{3}$m                        & a = 6.0602(4) \AA \\ [1ex]
\hline
\end{tabular}
\end{adjustbox}
\end{table*}

\begin{figure}[h]
\begin{center}
\includegraphics[width=\columnwidth]{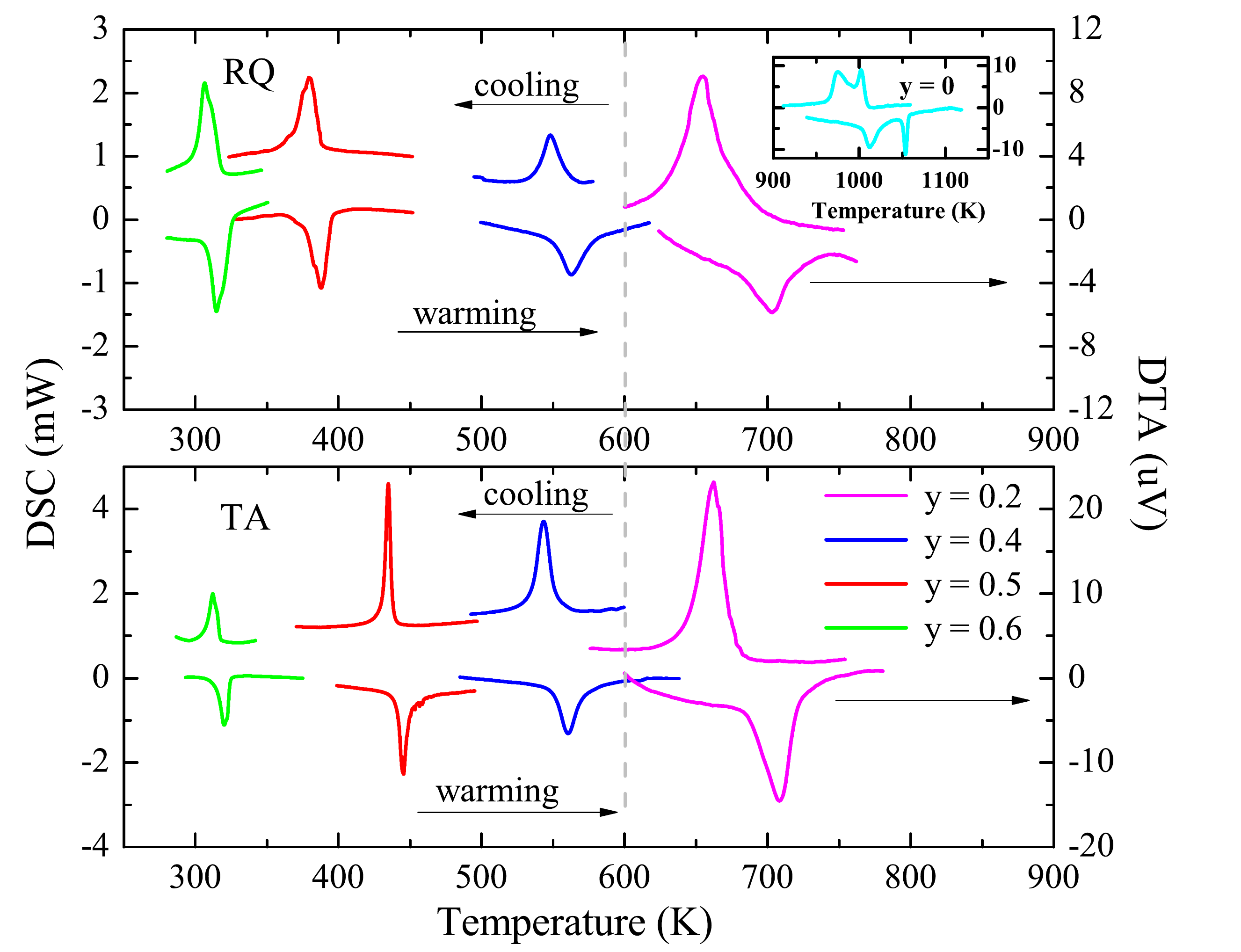}
\caption{The thermo-analytical measurements for the rapid quenched (RQ) and temper annealed (TA) alloys showing DSC and DTA plots as a function of temperature for 0.6 $\geq y \geq$ 0.4 and $y$ = 0.2 respectively. The DTA scan for the alloy $y$ = 0 is shown as an inset.}
\label{fig:DSC}
\end{center}
\end{figure}

\begin{figure}[h]
\begin{center}
\includegraphics[width=\columnwidth]{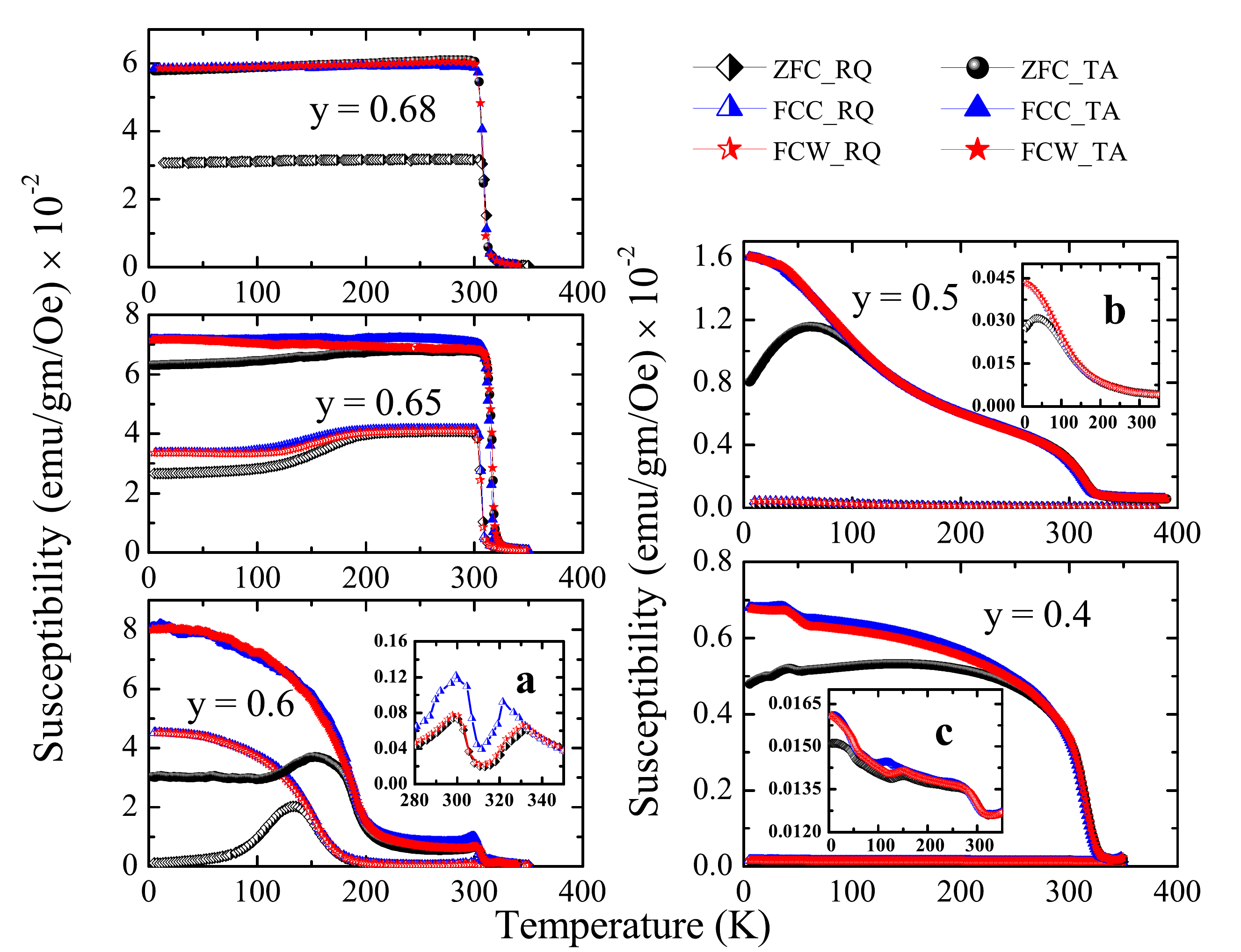}
\caption{Susceptibility as a function of temperature for the rapid quenched (RQ) alloys 0.68 $\geq y \geq$ 0.4 and their temper annealed (TA) counterparts. Inset (a) features the martensitic transition in the RQ $y$ = 0.6 while insets (b) and (c) present the susceptibility curves of RQ $y$ = 0.5 and $y$ = 0.4 alloys respectively in an amplified scale.}
\label{fig:MT}
\end{center}
\end{figure}

Differential scanning calorimetry (DSC) (273 K $\le T \le$ 850 K) and Differential thermal analysis (DTA) (300 K $\le T \le$ 1100 K) measurements were performed on the transforming alloys 0.2 $\leq y \leq$ 0.6 to determine the martensitic transition temperature of the RQ and their TA counterparts. The results are presented in Fig. \ref{fig:DSC}. The inset shows the DTA plot of the alloy NiMn displaying martensitic transition at around 990 K and $L2_1$ to $B_2$ ordering transition at about 1032 K. The expected decrease in the martensitic transition temperature of RQ alloys is seen with increase in In content from $y$ = 0 to $y$ = 0.6. Interestingly, none of the TA alloys, from $y$ = 0.6 to the completely disintegrated $y$ = 0.2, exhibit any significant change in the transformation temperature.

Temperature dependent magnetization results for the RQ and TA alloys are presented in Fig. \ref{fig:MT}. An increase in the magnetic moment seen in all alloy compositions upon temper annealing, hints at advancement in ferromagnetism. The $y$ = 0.68 and $y$ = 0.65 RQ alloys are ferromagnetic with an ordering temperature, $T_C$ = 310 K. Upon temper annealing, the $T_C$ increases slightly to 320 K in the $y$ = 0.65 alloy while it remains nearly constant in the alloy $y$ = 0.68. The $y$ = 0.6 alloy transforms below 333 K (inset (a) highlights the martensitic transition) with a magnetically glassy ground state comprising of ferromagnetic and antiferromagnetic clusters \cite{Nevgi322020}. On temper annealing, the alloy displays two ferromagnetic transitions at $T_C$ = 320 K and $T_C$ = 200 K. A complete change of magnetic character is seen in the case of $y$ = 0.5 and 0.4. Both the RQ alloys, $y$ = 0.5 and 0.4 transform from an antiferromagnetic state (insets (b) and (c) respectively) to a state with dominant ferromagnetic interactions as indicated by a $T_C$ = 320 K. An increase in ferromagnetic behavior is an indication of the growth of $L2_1$ phase, which seems apparent in all the TA alloys.

The experiments so far have shown a varying degree of structural degradation ranging from total phase separation in $y$ = 0.2 to only subtle structural changes like an increase in the width of the Bragg peaks ($y$ = 0.65 and 0.68) post temper annealing. Despite these varying degree of structural changes, the martensitic transition temperatures remain nearly the same in all the transforming alloy compositions. The magnetization measurements also display an upsurge in ferromagnetism in all the alloy compositions after temper annealing. The strengthening of ferromagnetism could be a signature of segregation of Ni$_2$MnIn Heusler phase. However, the nearly invariant martensitic transition temperature, even in a completely phase separated Ni$_2$Mn$_{1.8}$In$_{0.2}$ is puzzling. Such a scenario is possible if the decrease in martensitic transition temperature of Ni$_2$Mn$_{2-y}$In$_y$  with increasing $y$ is ascribed to the dilution of NiMn entities by Heusler Ni$_2$MnIn structural units. The progressive replacement of Mn by In in Ni$_2$Mn$_{2-y}$In$_y$ results in the formation of Heusler structural entities in the NiMn matrix and, therefore, a randomly quenched off stoichiometric composition of Ni$_2$Mn$_{2-y}$In$_y$ would consist of a random distribution of $L2_1$ (Ni$_2$MnIn) and $L1_0$ (NiMn) structural units packed within a single crystal structure. However, at a local structural level, the presence of two structural entities should be visible. To explore such a possibility, room temperature Ni K and Mn K EXAFS data were analyzed and compared for both RQ and TA alloys of Ni$_2$Mn$_{2-y}$In$_y$.

\begin{figure}[h]
\begin{center}
\includegraphics[width=\columnwidth]{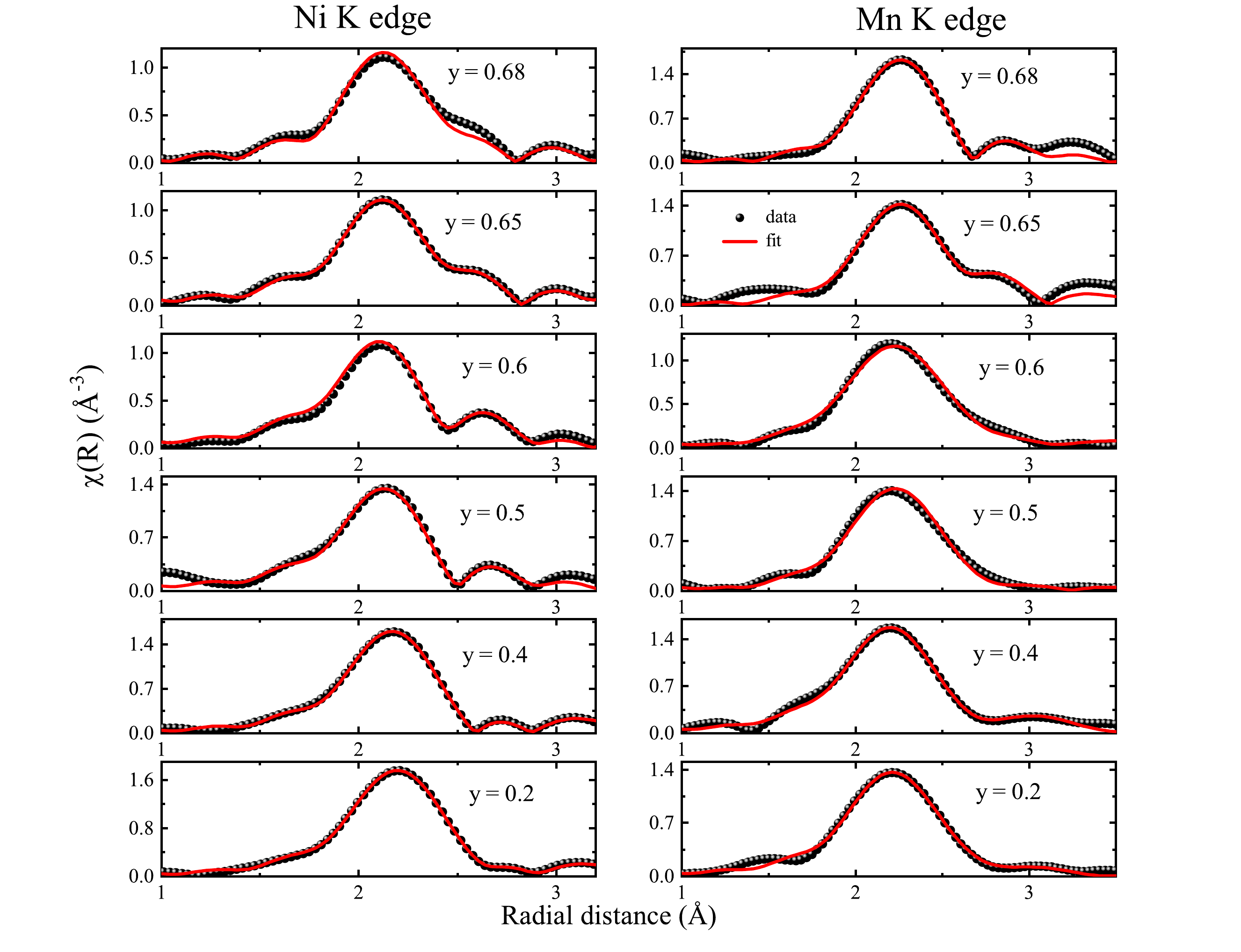}
\caption{The magnitude of Fourier transform spectra at Ni K and Mn K edges in rapid quenched (RQ) alloy compositions in the series Ni$_2$Mn$_{2-y}$In$_{y}$ at 300 K.}
\label{fig:EXAFS_AP}
\end{center}
\end{figure}

\begin{figure}[h]
\begin{center}
\includegraphics[width=\columnwidth]{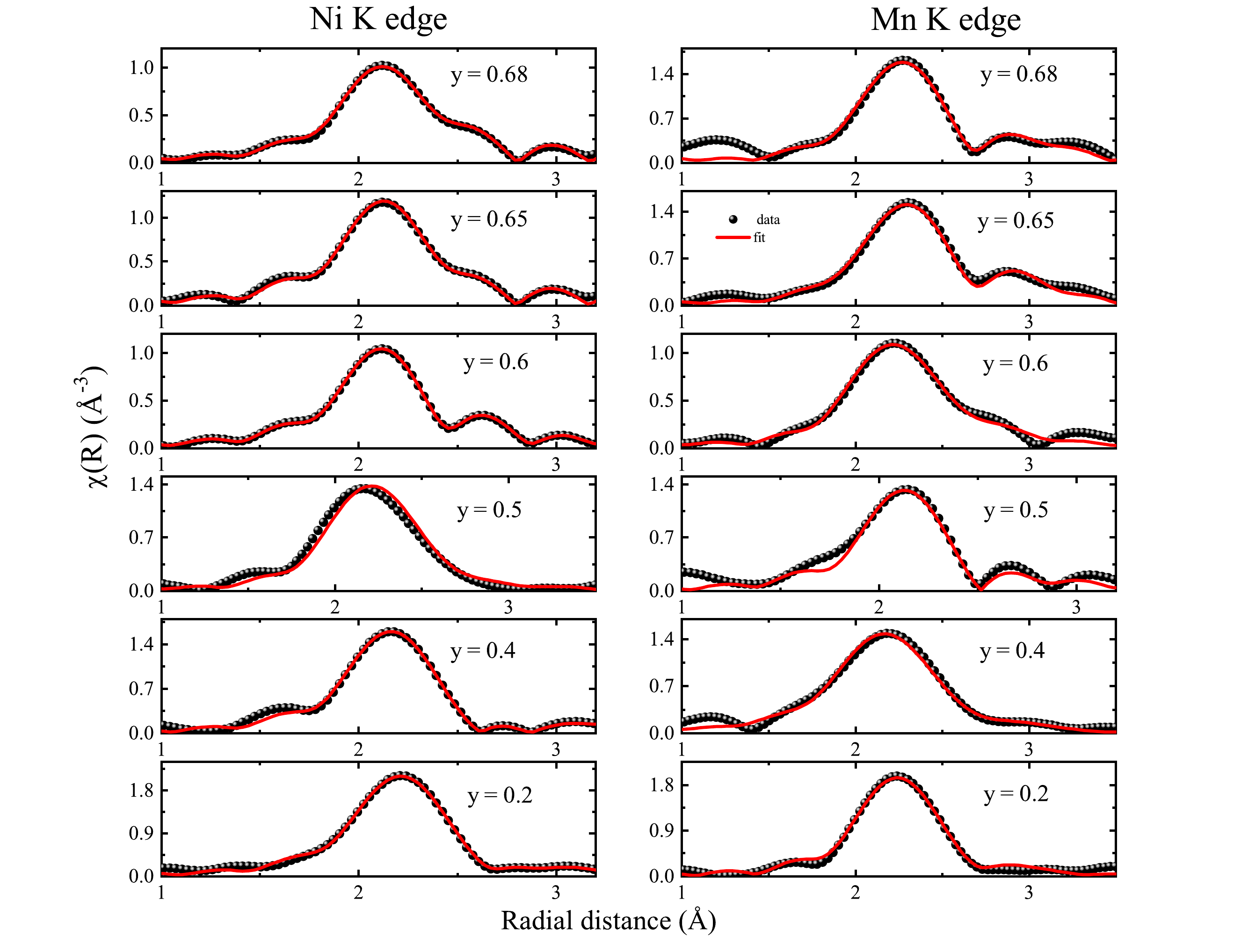}
\caption{The magnitude of Fourier transform spectra at Ni K and Mn K edges in temper annealed (TA) alloy compositions in the series Ni$_2$Mn$_{2-y}$In$_{y}$ at 300 K.}
\label{fig:EXAFS_TA}
\end{center}
\end{figure}

The Ni K and Mn K edge EXAFS spectra of Ni$_2$MnIn ($y$ = 1) and NiMn ($y$ = 0) were fitted using the correlations obtained from the respective cubic $L2_1$ and tetragonal $L1_0$ crystal structures. The structural model described in references \onlinecite{Nevgi322020,Bhobe200674} was employed to fit the EXAFS spectra in the off stoichiometric RQ alloys. On the other hand, the best fits to the EXAFS data in all the TA alloys were obtained using a structural model consisting of correlations from both $L2_1$ and $L1_0$ structures. Accordingly, to fit the Ni K edge EXAFS spectra, three correlations from the $L2_1$ structure arising due to the nearest neighbors, Mn and In at $\sim$ 2.6 \AA, and the next nearest Ni atoms at $\sim$ 3 \AA~ along with three correlations, Ni--Mn at $\sim$ 2.5 \AA~ and Ni--Ni at $\sim$ 2.6 \AA, and $\sim$ 3.5 \AA~ from the $L1_0$ structure were considered. Likewise, the Mn K edge EXAFS spectra were fitted considering Mn--Ni correlation at $\sim$ 2.6 \AA~ and Mn--In ($\sim$ 3 \AA) and Mn--Mn$_Z$ structural correlation arising from the replacement of In by Mn in the Z sublattice of the X$_2$YZ Heusler structure along with Mn--Ni ($\sim$ 2.5 \AA) and Mn--Mn ($\sim$ 2.6 \AA~ and $\sim$ 3.5 \AA) correlations from $L1_0$ structure. To improve the reliability of the fitting, both Ni K and Mn K EXAFS spectra were fitted together using the structural constraints imposed by $L1_0$ and $L2_1$ symmetry. All the fittings were carried out  in the $k$ range of 3 -- 12 \AA$^{-1}$ and $R$ range of 1 -- 3 \AA. A total of 17 independent parameters consisting of correction to the bond length $\Delta R$ and mean square variation in bond length $\sigma^2$ for each of the scattering paths were employed in fitting EXAFS spectra in TA and RQ alloys, respectively. The amplitude reduction factor, $S_{0}^2$ and correction to edge energy $\Delta E_0$, were obtained from the analysis of the standard metal spectra and were kept fixed throughout the analysis. The results of these fittings in the $R$ space, for both RQ and TA alloys, are presented in Figs. \ref{fig:EXAFS_AP} and \ref{fig:EXAFS_TA}, respectively.

\begin{figure}[h]
\begin{center}
\includegraphics[width=\columnwidth]{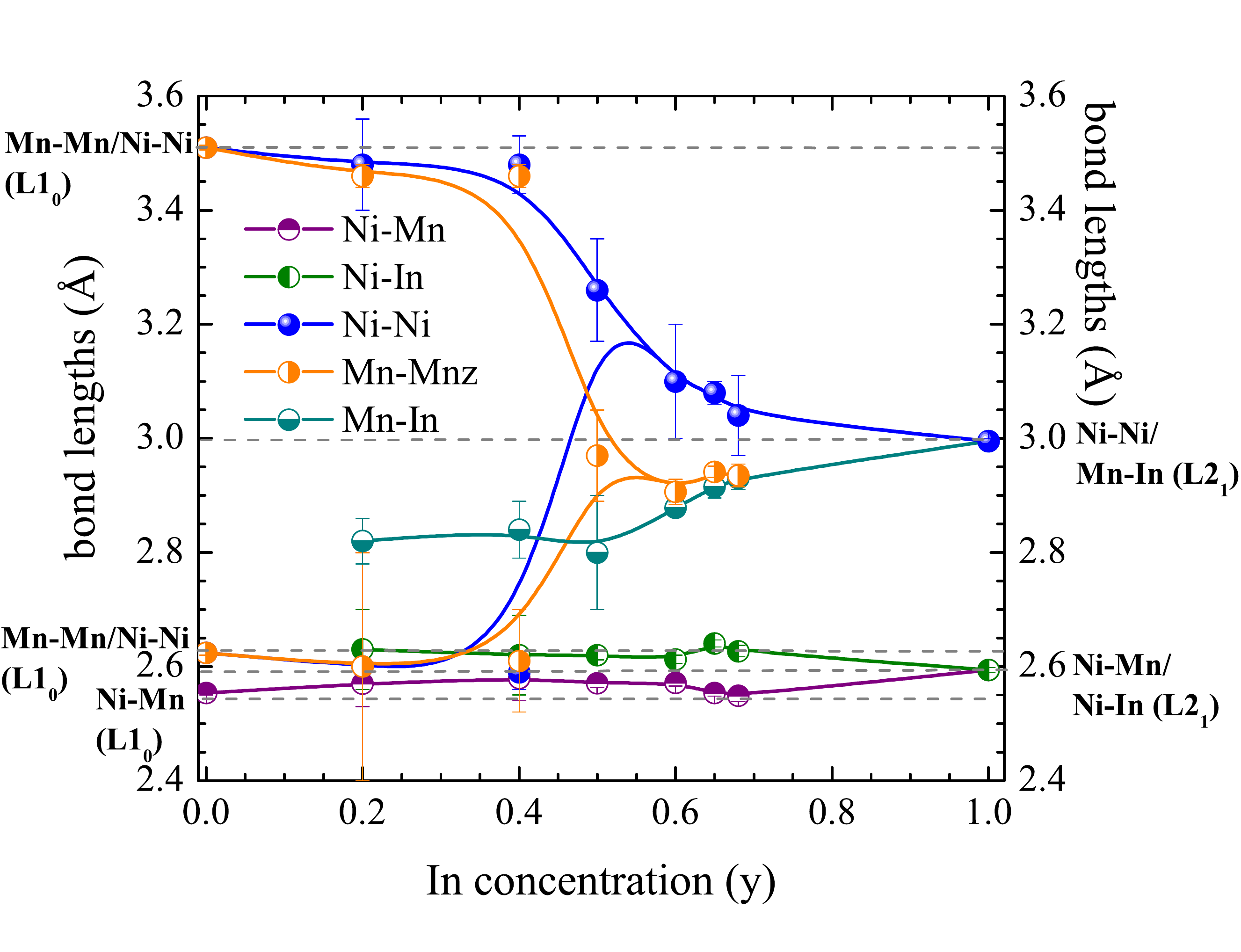}
\caption{Variation of bond lengths with In concentration in rapid quenched (RQ) Ni$_2$Mn$_{2-y}$In$_{y}$.}
\label{fig:EXAFS1}
\end{center}
\end{figure}

\begin{figure}[h]
\begin{center}
\includegraphics[width=\columnwidth]{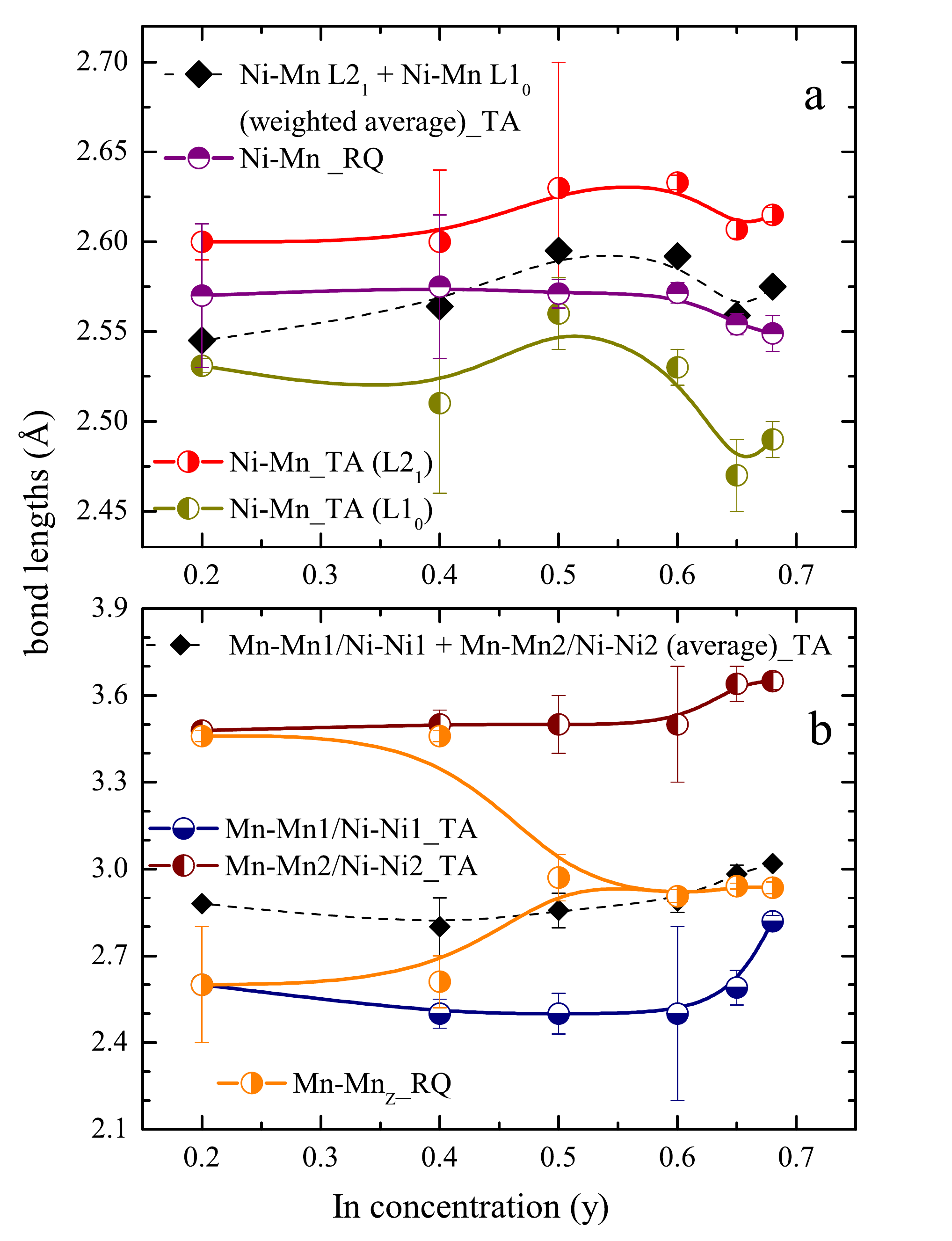}
\caption{Variation of selected nearest neighbor and next nearest neighbor bond lengths with In concentration in rapid quenched (RQ) and temper annealed (TA) Ni$_2$Mn$_{2-y}$In$_{y}$.}
\label{fig:EXAFS2}
\end{center}
\end{figure}

The variation of Ni--Mn, Ni--In, Ni--Ni, Mn--In and Mn--Mn$_Z$ bond distances as a function of In concentration in RQ alloys of the Ni$_2$Mn$_{2-y}$In$_{y}$ series is presented in Fig. \ref{fig:EXAFS1}. It can be seen that with decreasing In content, the Ni--Mn and Ni--In bond distances, which are supposed to be equal in an $L2_1$ Heusler structure, deviate away from each other with Ni--Mn bond length being shorter than Ni--In. With $y$ approaching zero, the Ni--Mn distance approaches a value of 2.51 \AA~, which is equal to Ni--Mn distance in tetragonal NiMn. Further, with the appearance of Mn--Mn$_Z$ bond, the otherwise equal Ni--Ni and Mn--In bond lengths also diverge away from their lattice calculated value of $\sim$ 3 \AA. While the Ni--Ni bond distance increases and eventually splits into two bonds at about $y$ = 0.5. The longer Ni--Ni bond increases to 3.51 \AA~ while the shorter Ni--Ni bond converges to 2.61 \AA. Both these values of bond distances are equal to Ni--Ni bonds in the tetragonal phase of NiMn. Similar behavior is also exhibited by the Mn--Mn$_Z$ distance. On the other hand, the Mn--In bond distance decreases slowly and saturates at $\sim$ 2.8 \AA~  for $y = 0.2$.

Fig. \ref{fig:EXAFS2} (a) compares the variation of nearest neighbor Ni--Mn bond distances in the off stoichiometric TA and RQ Ni$_2$Mn$_{2-y}$In$_y$ alloys. There are two Ni-Mn bond distances in the TA alloys arising from Ni$_2$MnIn and NiMn structural units. Their values agree quite well with those calculated from their respective crystal structures. Further, it can be seen that the weighted average of these two Ni--Mn bond lengths nearly equals the Ni--Mn bond length in the RQ alloys directly indicating the formation of NiMn and Ni$_2$MnIn type structural entities in the rapidly quenched off stoichiometric Ni$_2$Mn$_{2-y}$In$_y$ alloys.

The comparison of second nearest neighbor Mn--Mn$_Z$ distance in the RQ Ni$_2$Mn$_{2-y}$In$_y$ alloys and their TA counterparts is shown in Fig. \ref{fig:EXAFS2} (b). In the non martensitic RQ compositions ($y \geq 0.65$), Mn--Mn$_Z$ as well as Mn--In and Ni--Ni (see Fig. \ref{fig:EXAFS1}) are equal to about 3 \AA~, which matches with the values obtained from the unit cell parameters. Therefore, in this range $1 > y \geq 0.65$, the structural distortions due to the formation of NiMn and Ni$_2$MnIn structural entities, are limited only up to the nearest neighbor. In the case of TA alloys, the two Mn--Mn distances also appear to converge towards each other as $y$ increases to 1. For the RQ martensitic alloys with a modulated structure ($0.6 \ge y > 0.4$), the Mn--Mn$_Z$ distance increases before splitting into two and eventually equalizes with Mn--Mn bond distances in the TA alloys for all $y \le 0.4$. A similar behavior is also displayed by the second neighbor Ni--Ni bonds. Interestingly, the average of the two Mn--Mn bond distances in the TA alloys is equal to the value of Mn--In distance obtained in RQ alloys right through the series.

\section{Discussion}

\begin{figure}[h]
\begin{center}
\includegraphics[width=\columnwidth]{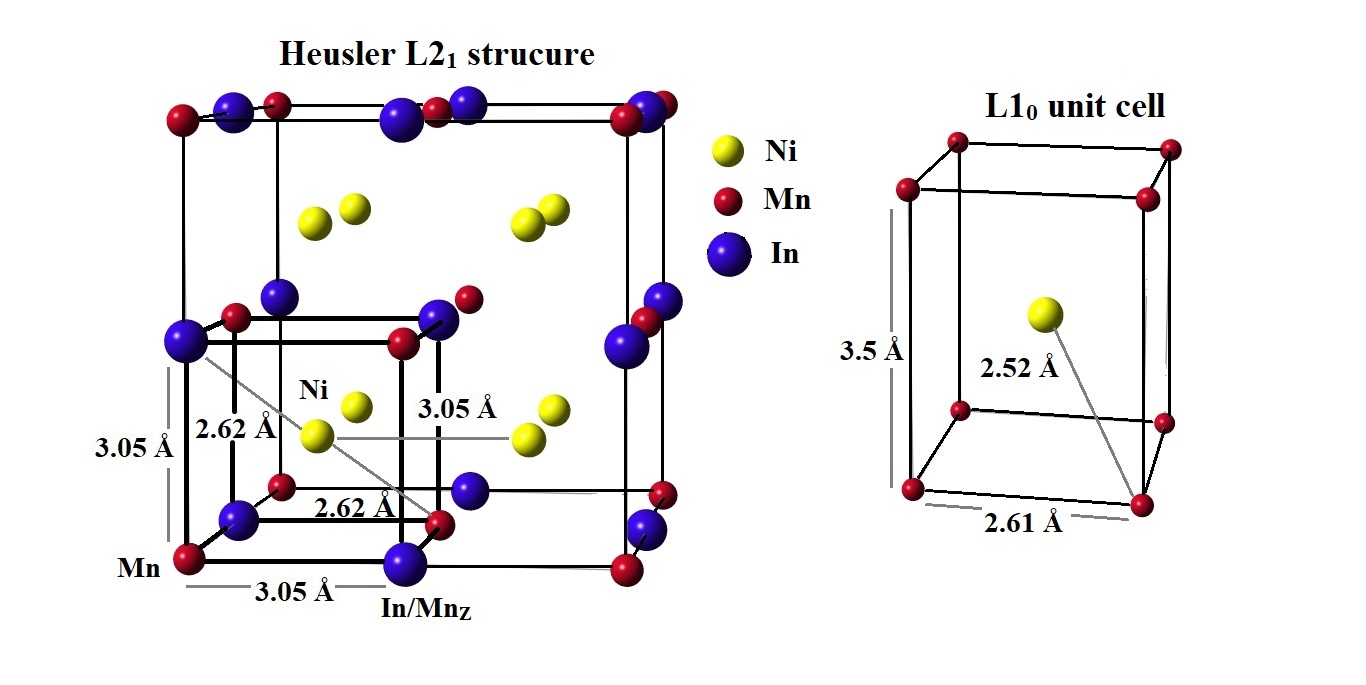}
\caption{The crystallographic model for the L2$_1$ unit cell of Ni$_2$MnIn and L1$_0$ unit cell of NiMn. The Mn$_Z$ represent the Mn atom in place of In site in the Heusler structure.}
\label{fig:structure}
\end{center}
\end{figure}

Doping impurity atoms like In in the binary NiMn alloy transforms its martensitic $L1_0$ structure to austenitic cubic $L2_1$ Heusler structure. The magnetic ground state also changes from antiferromagnetic to ferromagnetic. The transition occurs beyond a critical value of In concentration ($y = 0.65$) without appearance of non-ergodic phases like strain glass \cite{Nevgi322020}. The transition from $L1_0$ to $L2_1$ structure occurs via an incommensurate modulated structure. Temper annealing of any of the off stoichiometric compositions of Ni$_2$Mn$_{2-y}$In$_y$ is reported to disintegrate into y[Ni$_2$MnIn] + (1-y)[(NiMn)$_2$] \cite{Cakir1272017}.

Schematic representation of the two structures is shown in Fig. \ref{fig:structure}. The tetragonal $L1_0$ structure consists of nearest neighbor Ni--Mn bond distance of 2.52 \AA~ and next nearest neighbor Ni--Ni/Mn--Mn bond distances at 2.61 \AA~ and 3.51 \AA. On the other hand, the cubic $L2_1$ superstructure of Ni$_2$MnIn, formed due to rocksalt type ordering of Mn and In atoms, consists of equidistant Ni--Mn and Ni--In nearest neighbors at $\sim$ 2.6 \AA~ and next nearest Ni--Ni and Mn--In bonds at about 3.0 \AA. In Ni$_2$Mn$_{2-y}$In$_y$ additional Mn--Mn$_Z$ bonds at about 3 \AA~ develop due to partial occupation of Z sites by Mn and In atoms in X$_2$YZ Heusler structure.

EXAFS analysis indicate the Ni--Mn bond distance to be always shorter than Ni--In bond distance in all the off stoichiometric, including cubic ($y \ge 0.65$) Ni$_2$Mn$_{2-y}$In$_y$ alloys. In temper annealed alloys, there are two nearest neighbor Ni--Mn distances. One due to the $L2_1$ component and the other due to $L1_0$ structural component. Interestingly, as seen from Fig. \ref{fig:EXAFS2} (a), the weighted average of the two nearest neighbor Ni--Mn distances in the TA alloys is nearly equal to the Ni--Mn bond distance obtained from EXAFS studies of RQ alloys. Thus, the shorter Ni--Mn distance as compared to Ni--In bond distance, observed in all rapidly quenched off stoichiometric compositions is a result of  proportionate addition of Ni--Mn distance in Ni$_2$MnIn and Ni--Mn distance in NiMn.  Thus the $L2_1$ and $L1_0$ structural entities develop in all the off stoichiometric RQ alloys. However, in the rapidly quenched alloys with $y \geq 0.65$, the structural distortions are limited only to the nearest neighbor. This is perhaps because, the phase fraction of $L1_0$ (NiMn) type entities in these alloys is small and their tetragonal distortion is accommodated within the $L2_1$ cubic supercell. A similar but reverse effect is observed in alloys with $y < 0.6$. In this case the average structure obtained from diffraction is tetragonal but with broad Bragg peaks. The diffraction peaks narrow down upon phase separation post temper annealing \cite{Dincklage82018}.

In the RQ alloys undergoing martensitic transformation ($0.6 \ge y > 0.4$), an increase in the phase fraction of $L1_0$ entities results in an extension of structural distortions to the second neighbor bond distances. For instance, the Ni--Ni and Mn--Mn$_Z$ distances deviate away from their crystallographic values and then split into two components as expected for tetragonal symmetry. On the other hand, the Mn--In bond distance, decreases to 2.8 \AA~ and remains nearly constant with decreasing In content. Even the third neighbor Mn--Mn distance also increases from 4.2 \AA~ to 4.35 \AA~ weakening the ferromagnetic order \cite{Nevgi322020}. Presence of structural distortions to second neighbor correlations results in a transition from $L2_1$ cubic structure to a modulated structure.

Ni$_2$Mn$_{1.35}$In$_{0.65}$ which has a biphasic structure with 88\% cubic and 12\% modulated phases, is the critical composition beyond which the martensitic transition disappears. The structural distortions in the second neighbor correlations also vanish for all higher values of In doping. Further, the value of Mn--In bond distance tracks the average value of the two Mn--Mn distances obtained for TA alloys. Therefore, it appears that the modulated crystal structure of the martensitic Ni$_2$Mn$_{2-y}$In$_y$ alloys is made up of randomly arranged Ni$_2$MnIn and NiMn type structural entities at a local structural level. The Mn--In bonds act as the bridging bonds compensating for the difference in the near neighbor bond distances of cubic Ni$_2$MnIn and tetragonal NiMn. Such a random arrangement of Ni$_2$MnIn and NiMn type structural entities within the crystal structure of the rapidly quenched Ni$_2$Mn$_{2-y}$In$_y$ alloys prevents the segregation of a defect phase which supports the absence of the transition from the martensitic ground state to strain glass in these alloys. On the other hand, the same random arrangement of ferromagnetic Ni$_2$MnIn and antiferromagnetic NiMn structural entities in the rapidly quenched alloys facilitates strongly interacting non-ergodic magnetic ground states like super spin glass. 

\section{Conclusions}

The present study on rapidly quenched and temper annealed Ni$_2$Mn$_{2-y}$In$_y$ alloys reveals presence of randomly packed Ni$_2$MnIn and NiMn structural units in all off stoichiometric alloys. Further, despite local structural disorder extending up to second nearest neighbor and metastable crystal structure, absence of a non-ergodic state like the strain glass remains intriguing rapidly quenched alloys. These units segregate and phase separate into proportionate amounts of $L2_1$ and $L1_0$ phases upon temper annealing. The random arrangement of $L2_1$ and $L1_0$ structural entities within a single crystal structure precludes segregation of point defects and thus preventing the system from exhibiting non-ergodic elastic behavior like strain glass. The random packing of the Ni$_2$MnIn and NiMn structural entities also seems to be responsible for the observed invariance of martensitic transition temperature between rapidly quenched and temper annealed alloys.

\section{Acknowledgements}

KRP and RN thank the Science and Engineering Research Board, Govt. of India under the project SB/S2/CMP-0096/2013 for financial assistance and Department of Science and Technology, Govt. of  India for the travel support within the framework of India\@ DESY collaboration. RN acknowledges the Council of Scientific and Industrial Research, Govt. of India for Senior Research fellowship. Edmund Welter and Ruidy Nemausat are thanked for experimental assistance at P65 beamline, PETRA III, DESY Hamburg. Some of the EXAFS experiments were also performed at BL-9C, Photon Factory, KEK, Japan as a part of proposal No. 2016G0132. Prof. A. K. Nigam is thanked for useful discussions.

\bibliographystyle{apsrev4-1}
\bibliography{Ref}

\end{document}